\documentclass[prd,aps,floatfix,12pt]{revtex4}

\usepackage{psfig}
\usepackage{verbatim}
\usepackage{color}
\usepackage{rotate}
\usepackage{graphics}
\usepackage{epsf}
\usepackage{epsfig}
\usepackage{graphicx}
\usepackage{latexsym}
\usepackage{amscd}
\usepackage{amssymb}
\newcommand{\beq}{\begin{equation}}
\newcommand{\eeq}{\end{equation}}
\newcommand{\beqn}{\begin{eqnarray}}
\newcommand{\eeqn}{\end{eqnarray}}

\newcommand\eps\varepsilon
\newcommand\euler{{\rm e}}
\newcommand\imag{{\rm i}}
\newcommand\kzero{{\rm K}_0}
\newcommand\kone{{\rm K}_1}

\def\GeV{\,\mbox{GeV}}
\def\TeV{\,\mbox{TeV}}
\def\lsim{\mathrel{\rlap{\lower4pt\hbox{\hskip1pt$\sim$}}
    \raise1pt\hbox{$<$}}}         
\def\gsim{\mathrel{\rlap{\lower4pt\hbox{\hskip1pt$\sim$}}
    \raise1pt\hbox{$>$}}}         

\begin{document}

\hfill {LA-UR-02-6304}
\title{Relating parton model and color dipole formulation of\\ heavy
quark hadroproduction}
\author{
J\"org Raufeisen$^a$ and
Jen-Chieh Peng$^b$}
\affiliation{$^a$ Los Alamos National Laboratory, MS H846, Los Alamos,
		  NM 87545, USA\\
	     $^b$ Department of Physics, University of Illinois, Urbana, 
		IL 61801, USA
		}


\vspace*{2cm}

\begin{abstract}
\vspace*{1cm}
\centerline{\bf Abstract}
\noindent
At high center of mass energies, hadroproduction of heavy quarks can 
be expressed in terms of the same color dipole cross section as low
Bjorken-$x$ deep inelastic scattering. We show analytically that
at leading order,
the dipole formulation is equivalent to the 
gluon-gluon fusion mechanism of the
conventional parton model.
In phenomenological application, we employ a parameterization of the dipole
cross section which also includes higher order and saturation effects,
thereby going beyond the parton model. Numerical calculations
in the dipole approach agree well with experimental data on open charm 
production over a wide range of energy. Dipole approach and 
next to leading order parton model yield similar values for open
charm production, but for open bottom production, the dipole approach tends
to predict somewhat higher cross sections than the parton model. 
\smallskip

\end{abstract}
\maketitle    

\clearpage

\section{Introduction}\label{sec:intro}

Heavy quark hadroproduction has been conventionally described in the
framework of QCD parton model \cite{pm1,pm2,pm3}. 
At high center of mass energies 
$\sqrt{s}$, many hard-processes, including Drell-Yan \cite{boris} and 
heavy-quark production \cite{npz,kt}, can be described in terms of the 
color dipole cross sections originally deduced from low $x_{Bj}$ 
deep inelastic scattering (DIS) (see {\em e.g.} \cite{gBFKL}). 
The dipole formulation of 
heavy quark production was first introduced in \cite{npz}. This
alternative approach to heavy quark production provides a theoretical
framework for treating the nuclear effects, which are present in high
energy proton-nucleus and nucleus-nucleus collisions. However, the 
connection between this dipole approach and the conventional parton model
approach remains to be delineated and clarified. The purpose of this paper
is to demonstrate the validity of the dipole approach in proton-proton
($pp$) collisions and to illuminate its relation to the conventional parton
model.

An alternative approach to heavy quark production that is designed especially
for energies much larger than the heavy quark mass $m_Q$ and which
is able to describe nuclear effects is desirable for a
variety of reasons.
At low $x$, the heavy quark pair is produced over large
longitudinal distances, which can exceed the radius of a large nucleus
by orders of magnitude. 
Indeed, even though the matrix element of a hard 
process is dominated by short distances, of the order of the
inverse of the hard scale, the cross section of that process also depends
on the phase space element. 
Due to gluon radiation, the latter becomes very large at high energies,
and it is still a challenge how to resum the corresponding low-$x$ logarithms.
The dipole formulation allows for a 
simple phenomenological recipe to include these low-$x$ logs.
The large length scale in the problem 
leads to pronounced nuclear effects, giving one the possibility to
use nuclear targets as microscopic detectors to study the space-time 
evolution of heavy quark production. 

In addition, heavy quark 
production is of particular interest, because this
process directly probes the gluon distributions of the colliding particles.
Note that at the tremendous center of mass energies of the Relativistic
Heavy Ion Collider
(RHIC) and especially of the Large Hadron Collider (LHC), 
charm (and at LHC also bottom) decays 
will dominate the dilepton continuum \cite{gmrv}. Thus,
a measurement of the heavy quark production cross section at RHIC and LHC
will be relatively easy to accomplish
and can yield invaluable information 
about the (nuclear) gluon density \cite{ev}.
It is expected that at very low $x$, 
the growth of the gluon density will be slowed 
down by nonlinear terms  
in the QCD evolution equations \cite{glr}. The onset of this
non-linear regime is controlled by the so-called saturation scale $Q_s(x,A)$,
which is already of 
order of the charm quark mass at RHIC and LHC energies.
Moreover, $Q_s(x,A)\propto A^{1/3}$ ($A$ is the atomic mass of the nucleus), 
so that 
one can expect sizable higher twist corrections in $AA$ collisions \cite{kt}. 
Note that saturation will lead to a breakdown of the
twist expansion, since 
one cannot conclude any more that terms suppressed by powers of the 
heavy quark mass $m_Q$ are small, $Q_s^n(x,A)/m_Q^n\in O(1)$ for any $n$.
Saturation effects are most naturally described in the dipole 
picture.

\section{Color dipole approach to heavy quark hadroproduction}
\label{sec:dipole}

\begin{figure}[t]
  \centerline{\scalebox{0.65}{\includegraphics{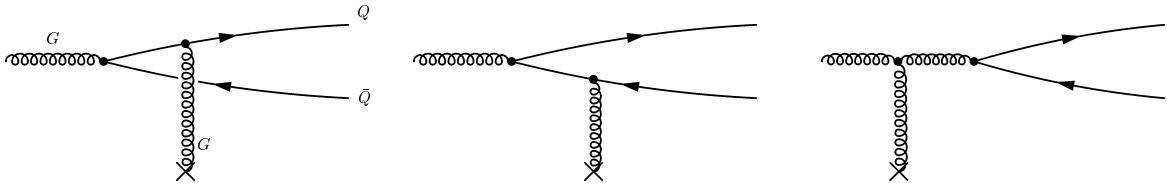}}}
    \center{\parbox[b]{13cm}{\caption{\em
      \label{fig:3graphs}The three lowest order
graphs contributing to heavy quark
 production in the dipole approach.  
These graphs correspond to the gluon-gluon
fusion mechanism of heavy quark production in the parton model.}  
    }  }
\end{figure}

The color dipole approach is formulated in the
target rest frame, where heavy quark production looks like pair creation
in the target color field, Fig.~\ref{fig:3graphs}.
For a short time,
a gluon $G$ from the projectile hadron
can develop a fluctuation which contains a heavy quark
pair ($Q\bar Q$). Interaction with the 
color field of the target 
then may release these heavy quarks. The similarity between
heavy quark production and pair creation at high 
partonic center of mass
energies has already 
been pointed out in \cite{pm1}. Apparently, the mechanism depicted in
Fig.~\ref{fig:3graphs} corresponds to the gluon-gluon
fusion mechanism of heavy quark production in the leading order (LO)
parton model.
The dipole formulation is therefore applicable only at low $x_2$, where
the gluon density of the target is much larger than all quark 
densities\footnote{We use standard kinematical
variables, $x_2=2P_{Q\bar Q}\cdot P_1/s$ and 
$x_1=2P_{Q\bar Q}\cdot P_2/s$,
where $P_1$ ($P_2$)
is the four-momentum of the projectile (target) hadron, and
$P_{Q\bar Q}$ is the four-momentum of the heavy quark pair.
In addition, $M_{Q\bar Q}$ is the invariant 
mass of the pair, and $s$ is the hadronic center of mass energy 
squared.}. 
The kinematical range where the dipole approach is valid can of 
course only be determined a posteriori. This is similar to determining the
minimal value of $Q^2$ for which perturbative QCD still works.

We will now present formulae for the partonic cross section for producing
a heavy quark pair $Q\bar Q$, limiting ourselves 
to presenting only the final result. The production amplitudes can be found in
\cite{kt}.
The derivation can be performed along the same lines as for Drell-Yan,
which is explained in detail in the appendix of \cite{rpn}.

After summation over all three color states in which the 
$Q\bar Q$ pair in Fig.~\ref{fig:3graphs} can be produced, one obtains for the
partonic cross section \cite{kt},
\beq\label{eq:all}
\sigma(GN\to \{Q\bar Q\} X)
=\int_0^1 d\alpha \int d^2\rho 
\left|\Psi_{G\to Q\bar Q}(\alpha,\rho)\right|^2
\sigma_{q\bar q G}(\alpha,\rho),
\eeq 
where $\sigma_{q\bar qG}$ is
the cross section for scattering a color neutral quark-antiquark-gluon
system on a nucleon \cite{kt},
\beq\label{eq:qqG}
\sigma_{q\bar qG}(\alpha,\rho)
=\frac{9}{8}\left[\sigma_{q\bar q}(\alpha\rho)
+\sigma_{q\bar q}({\bar\alpha}\rho)\right]
-\frac{1}{8}\sigma_{q\bar q}(\rho).
\eeq
Here $\alpha$ is the light-cone momentum fraction carried by the 
heavy quark $Q$,
and $\bar\alpha$
is the momentum fraction of the $\bar Q$. In LO, with no
additional gluon in the final state, $\alpha+\bar\alpha=1$.
The dipole cross section 
$\sigma_{q\bar q}(\rho)$ is an eigenvalue of the forward
diffraction amplitude
operator and has to be determined from experimental data. It depends on the 
transverse separation $\rho$ between quark and antiquark. 
We point out that 
$\sigma_{q\bar q}(\rho)$ is flavor independent, {\em i.e.} it is
the same for a dipole of heavy quarks ($Q\bar Q$) as for light quarks
($q\bar q$).
Note that
the dipole cross section would be independent of energy, 
if only the Born graphs in 
Fig.~\ref{fig:3graphs} were taken into account. 
However, higher order corrections will make
$\sigma_{q\bar q}$ a function $x_2$. 
In order to simplify the notation,
we do not explicitly write out the $x_2$ dependence of the dipole cross 
section.

The light-cone (LC) wavefunctions for the transition $G\to Q\bar Q$
can be calculated perturbatively,
\beqn\nonumber\label{eq:lcwf}
\Psi_{G\to Q\bar Q}(\alpha,\vec\rho_1)
\Psi^*_{G\to Q\bar Q}(\alpha,\vec\rho_2)&=&
\frac{\alpha_s(\mu_R)}{(2\pi)^2}\Biggl\{m_Q^2\kzero(m_Q\rho_1)\kzero(m_Q\rho_2)
\Biggr. \\
&+&\left.
\left[\alpha^2+{\bar\alpha}^2\right]m_Q^2
\frac{\vec\rho_1\cdot\vec\rho_2}{\rho_1\rho_2}
\kone(m_Q\rho_1)\kone(m_Q\rho_2)
\right\},
\eeqn
where $\alpha_s(\mu_R)$ is the strong coupling constant, which is probed
at a renormalization scale $\mu_R\sim m_Q$.
We work in a mixed representation, where the longitudinal direction is
treated in momentum space, while the transverse directions are described in
coordinate space representation. 

Partonic configurations with fixed transverse separations in
impact parameter space have been identified as eigenstates of the interaction 
a long time ago \cite{borisold,mp}. 
Since the degrees of freedom in the dipole approach
are eigenstates of the interaction, this approach is especially suitable to
describe multiple scattering effects, {\em i.e.} nuclear effects 
\cite{npz,kt}.

Eq.~(\ref{eq:all}) is a special case of the general rule that at high energy,
the cross section for the reaction $a+N\to\{b,c,\dots\}X$ can be expressed
as convolution of the LC wavefunction for the transition $a\to\{b,c,\dots\}$
and the cross section for scattering the color neutral
$\{{\rm anti-}a,b,c\dots\}$-system on the target nucleon $N$. 

Note that although 
the dipole cross section is flavor independent, the integral 
Eq.~(\ref{eq:all}) is not. Since the Bessel functions K$_{1,0}$ 
decay exponentially for large arguments, the largest values of $\rho$
which can contribute to the integral are of order $\sim 1/m_Q$. We
point out, that as a 
consequence of color transparency \cite{borisold,ct}, the dipole 
cross section vanishes $\propto\rho^2$ for small $\rho$. Therefore, the 
$Q\bar Q$ production cross section behaves roughly like $\propto 1/m_Q^2$
(modulo logs and saturation effects).

In order to calculate the cross section for heavy quark pair production
in $pp$ collisions,
Eq.~(\ref{eq:all}) has to be weighted with the projectile gluon density,
\beq\label{eq:dy}
\frac{d\sigma(pp\to \{Q\bar Q\}X)}{dy}
=x_1G\left(x_1,\mu_F\right)\sigma(GN\to \{Q\bar Q\} X),
\eeq
where $y=\frac{1}{2}\ln(x_1/x_2)$ is the
rapidity of the pair and $\mu_F\sim m_Q$.
In analogy to the parton model, we call $\mu_F$ the factorization scale. 
Uncertainties arising from the choice of this scale 
will be investigated in section 
\ref{sec:phenom}.
Integrating over all kinematically allowed rapidities yields
\beq\label{eq:total}
\sigma_{\rm tot}(pp\to \{Q\bar Q\}X)=2\int_0^{-\ln(\frac{2m_Q}{\sqrt{s}})}dy\,
x_1G\left(x_1,\mu_F\right)\sigma(GN\to\{Q\bar Q\} X).
\eeq

A word of caution is in order, regarding the limits of the 
$\alpha$-integration in Eq.~(\ref{eq:all}). Since the invariant mass of
the $Q\bar Q$-pair is given by
\beq\label{eq:invmass}
M^2_{Q\bar Q}=\frac{k_\perp^2+m_Q^2}{\alpha{\bar\alpha}},
\eeq
the endpoints of the $\alpha$-integration include configurations corresponding
to
arbitrarily large invariant masses, eventually exceeding the total available 
{\em cm.} energy. However, since $\rho$ and $k_\perp$ 
(the single quark transverse momentum) are conjugate variables, the pair mass
is not defined in the mixed representation, nor are the integration limits
for $\alpha$. Fortunately, this problem is present only at the very edge
of the phase space and therefore numerically negligible. 

Because of the mixed representation, in which the invariant mass of the pair
is not defined, the formulae for the $M_{Q\bar Q}$ distributions will 
be somewhat more complicated. In a first step, we present a formula for
the single quark $k_\perp$ distributions, which can be obtained after
a simple calculation from the amplitudes given in section 2.1 of 
\cite{kt}. One finds,
\beqn\label{eq:d2kall}
\nonumber
\frac{d^3\sigma(GN\to\{Q\bar Q\}X)}{d^2k_\perp d\alpha} &=&
\frac{1}{(2\pi)^2}\int d^2\rho_1 d^2\rho_2 
\euler^{\imag\vec k_\perp\cdot(\vec\rho_1-\vec\rho_2)}
\Psi_{G\to Q\bar Q}(\alpha,\vec\rho_1)
\Psi^*_{G\to Q\bar Q}(\alpha,\vec\rho_2)\\
\nonumber&\times&
\frac{1}{2}\Bigl\{\frac{9}{8}\left[
\sigma_{q\bar q}(\alpha\rho_1)
+\sigma_{q\bar q}(\bar\alpha\rho_1)
+\sigma_{q\bar q}(\alpha\rho_2)
+\sigma_{q\bar q}(\bar\alpha\rho_2)\right]
\Bigr.\\
&&\nonumber\qquad
-\frac{1}{8}\left[
\sigma_{q\bar q}(\alpha\vec\rho_1+\bar\alpha\vec\rho_2)
+\sigma_{q\bar q}(\bar\alpha\vec\rho_1+\alpha\vec\rho_2)\right]\\
&&\Bigl.\qquad
-\left[\sigma_{q\bar q}(\alpha|\vec\rho_1-\vec\rho_2|)
+\sigma_{q\bar q}(\bar\alpha|\vec\rho_1-\vec\rho_2|)\right]
\Bigr\}.
\eeqn
After integration over $k_\perp$, one obviously recovers
Eq.~(\ref{eq:all}).

The heavy quark pair invariant mass distribution is now easily obtained,
\beq\label{eq:dm2}
\frac{d\sigma(pp\to\{Q\bar Q\}X)}{d M_{Q\bar Q}^2}=
2\int_0^{-\ln(\sqrt{\tau})}dy\,
x_1G\left(x_1,\mu_F\right)
\int_{\alpha_{min}}^{\alpha_{max}}d\alpha\,
\alpha\bar\alpha\,\pi
\frac{d^3\sigma(GN\to\{Q\bar Q\}X)}{d^2k_\perp d\alpha},
\eeq
where $\tau=M^2_{Q\bar Q}/s$, and
the limits of the $\alpha$-integration now depend on $M_{Q\bar Q}$,
\beqn
\alpha_{max/min}&=&\frac{1}{2}\left(1\pm\sqrt{1-v}\right),\\
v&=&\frac{4m_Q^2}{M^2_{Q\bar Q}}.
\eeqn

It is possible to retrieve from Eq.~(\ref{eq:dm2}) the corresponding 
leading order parton model formula. Note that in leading order and
for small separations $\rho$, the dipole cross section can be expressed in 
terms of the target gluon density \cite{fsdipole},
\beq\label{eq:fsdipole}
\sigma_{q\bar q}(x,\rho)=\frac{\pi^2}{3}\rho^2\alpha_s(\mu)\,xG(x,\mu).
\eeq
In DIS, $\mu$ in Eq.~(\ref{eq:fsdipole}) is given by 
$\mu=\lambda/\rho$ where $\lambda$ is a number, since this is the only
available dimensionful scale at which the gluon density could be probed.
However, in the case of heavy quark production, it seems plausible
that $\mu$ is of order of $m_Q$, {\em i.e.} $\mu=\mu_{R,F}$.
Then, the curly bracket in Eq.~(\ref{eq:d2kall})
reduces to 
\beq
\Bigl\{\dots\Bigr\}_{Eq.~(\ref{eq:d2kall})}=
\frac{\pi^2}{3}\alpha_s(\mu_R)\,xG(x,\mu_F)
2\vec\rho_1\cdot\vec\rho_2
\left(\alpha^2-\frac{\alpha\bar\alpha}{4}+\bar\alpha^2\right),
\eeq
and 
it is possible to perform all but the 
$y$-integral in Eq.~(\ref{eq:dm2}) analytically. The result is
\beqn\label{eq:dm2ana}\nonumber
\frac{d\sigma(pp\to\{Q\bar Q\}X)}{d M_{Q\bar Q}^2}&=&
\alpha_s^2(\mu_R)\,
2\int_0^{-\ln(\sqrt{\tau})}dy\,
x_1G\left(x_1,\mu_F\right)
x_2G\left(x_2,\mu_F\right)\\
&\times&
\frac{\pi}{192M_{Q\bar Q}^4}
\left\{
\left(v^2+16v+16\right)
\ln\!\left(\frac{1+\beta}{1-\beta}\right)
-28\beta-31v\beta\right\},
\eeqn
where $\beta=\sqrt{1-v}$. Changing the integration variables to $x_1$
and $x_2$, we obtain for the total cross section,
\beqn\label{eq:totalana}\nonumber
\sigma_{\rm tot}(pp\to\{Q\bar Q\}X)&=&
\frac{\alpha^2_s(\mu_R)}{m_Q^2}\int dx_1 dx_2 G(x_1,\mu_F)G(x_2,\mu_F)\\
&\times&
\frac{\pi v}{192}\left\{
\left(v^2+16v+16\right)
\ln\!\left(\frac{1+\beta}{1-\beta}\right)
-28\beta-31v\beta\right\}.
\eeqn
Note that Eq.~(\ref{eq:totalana})
exactly agrees with the LO parton model result for the gluon-gluon 
contribution to the heavy quark production cross section 
(see Eqs.~(7,8,10,15) in \cite{pm1}). Thus, we have shown that
in leading order $\alpha_s$ and in leading twist approximation, 
dipole approach and parton model become 
equivalent at 
high energies, when the gluon-gluon contribution in the parton model 
dominates.
Furthermore, in leading-log $x_2$ approximation, 
the dipole cross section in the formulae for
heavy quark production is the same as in DIS, and is given by
Eq.~(\ref{eq:fsdipole}).
Note that Eqs.~(\ref{eq:dy}) and (\ref{eq:total}) are not analytically
equivalent to their parton model counterparts because of the inaccuracy in the
$\alpha$-integration. 

A calculation of higher order corrections is beyond the scope of this paper.
Nevertheless, higher order corrections are important, because they provide
a mechanism for the generation of large transverse momenta of the heavy quark 
pair. It would also be interesting to see, if the relation between 
$\sigma_{q\bar q}$ and $\sigma_{q\bar qG}$, Eq.~(\ref{eq:qqG}), persists 
to higher orders. It is possible to calculate higher order corrections
systematically in the dipole approach. This has been done in the case of DIS
in the generalized BFKL approach of Nikolaev and Zakharov, see {\em e.g.}
\cite{gBFKL}. However, the widely discussed next-to-leading order (NLO) 
correction to the BFKL equation \cite{nlo}, has left the
theory of low-$x$ resummation in an unclear state. We conclude this
section with the remark that, whatever the result of a higher order 
calculation will be, it will not be possible to reproduce the complete
NLO correction of the parton model \cite{pm1,pm2,pm3} in the dipole
approach. Only terms enhanced by a factor $\log(x_2)$ can be 
reproduced. This limitation is inherent to the dipole formulation.

\section{Phenomenological applications}\label{sec:phenom}

Now that it has been shown that the dipole formulation and the 
conventional parton model are equivalent in a certain approximation,
still the questions remain, how well does the dipole approach describe 
experimental data, and how much do predictions from both approaches
differ from each other. Note that we do not use the leading order gluon
density to calculate the dipole cross section according to 
Eq.~(\ref{eq:fsdipole}), instead we employ the phenomenological 
parameterization of \cite{bgbk} for 
$\sigma_{q\bar q}$, which reduces to Eq.~(\ref{eq:fsdipole})
in the limit $\rho\to 0$ and includes saturation effects at larger
transverse separations. This parameterization is an improved version of the 
saturation model presented in \cite{Wuesthoff1}, which now also includes 
DGLAP evolution. 
We use fit 1 of \cite{bgbk}, since the non-monotonous behavior
of fit 2 as a function of $\rho$ seems unphysical to us.
Both fits are constrained by HERA DIS data for $x_{Bj}\le0.01$.

We stress that $\sigma_{q\bar q}(x_2,\rho)$
contains much more information than the ordinary parameterizations of the
gluon density. The unintegrated gluon density is related to
the dipole cross section by Fourier transform 
(see {\em e.g} \cite{bgbk,Wuesthoff1}). Thus, 
$\sigma_{q\bar q}(x_2,\rho)$ also 
contains information about the transverse momentum distribution
of low-$x$ gluons in a nucleon.
The $x_2$ dependence and the intrinsic transverse momentum 
parameterized in the dipole cross section are higher order 
effects, which are taken into account in the dipole approach in 
this phenomenological way. Therefore, we compare 
predictions of the dipole approach for open charm and bottom
production to NLO
parton model calculation. The latter were performed with the code of
\cite{pm1,pm2,pm3}, using GRV98HO parton
distributions \cite{grv} in $\overline{\rm MS}$ scheme.  

In the dipole approach,
we use the one loop running coupling constant,
\beq
\alpha_s(\mu_R)=\frac{4\pi}
{\left(11-\frac{2}{3}N_f\right)\ln\!\left(\frac{\mu_R^2}{(200\,{\rm
MeV})^2}\right)}
\eeq 
at a renormalization scale $\mu_R\sim m_Q$, and
the number of light flavors is chosen to be $N_f=3$ for open charm
and $N_f=4$ for open bottom production. Furthermore, we use
the GRV98LO \cite{grv} gluon distribution (from CERNLIB \cite{cernlib})
to model the gluon density in the projectile. 
We use a leading order parton distribution function (PDF), because of
its probabilistic interpretation.
Note that one could
attempt to calculate the projectile gluon distribution from 
the dipole cross section. However, the projectile distribution functions
are needed mostly at large momentum fraction $x_1$, where the dipole cross
section is not constrained by data.  

\begin{figure}[t]
  \centerline{\scalebox{0.5}{\includegraphics{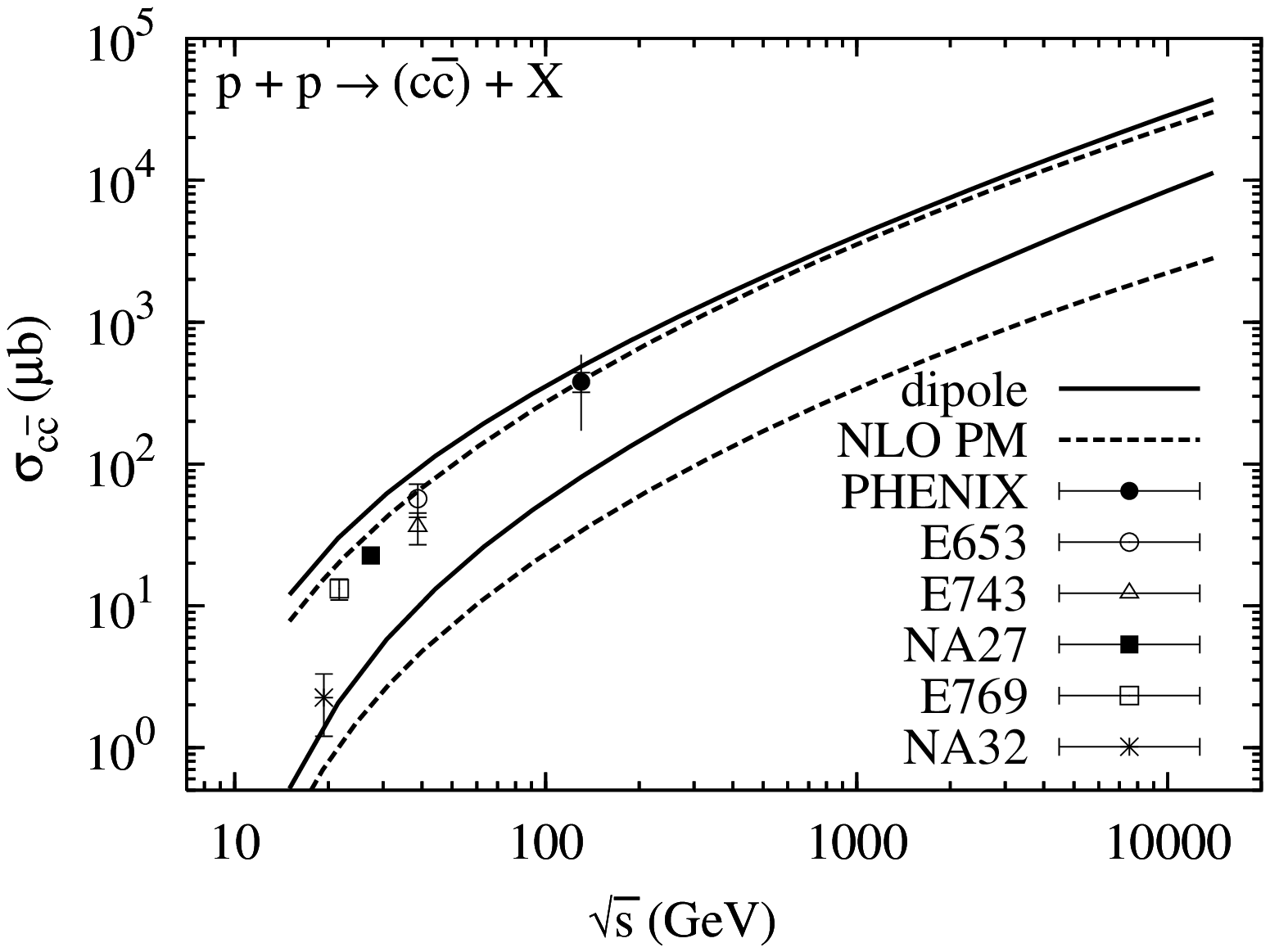}}
	      \scalebox{0.5}{\includegraphics{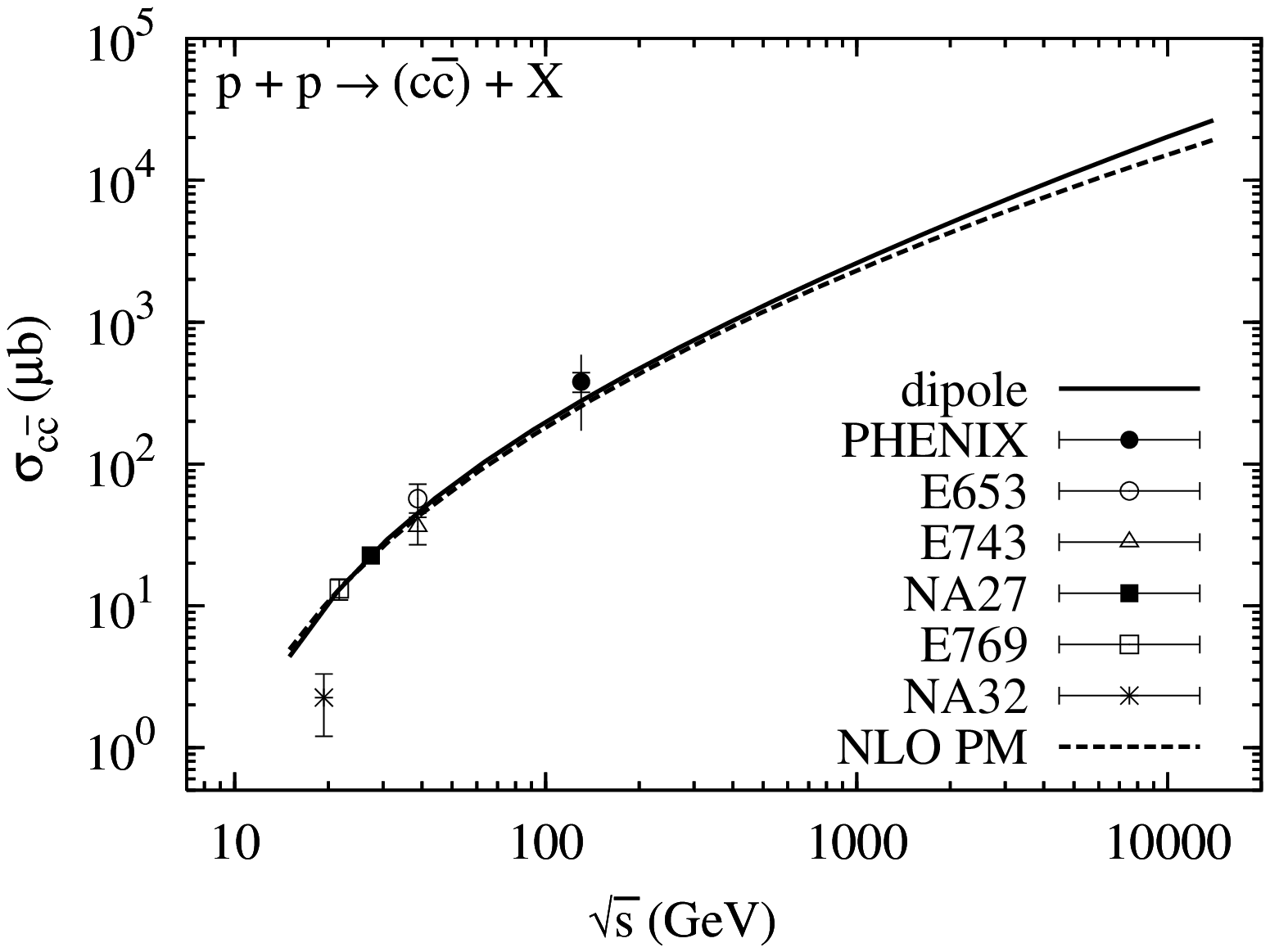}}}
\center{\parbox[thb]{13cm}{\caption{\em
      \label{fig:tcharm}
Results for the total open charm pair cross section as function 
of cm.\ energy. Varying free parameters in dipole approach (solid lines)
and in the 
parton model (dashed lines)
gives rise to the uncertainties shown on the left. In the
figure on the right, 
parameters in both models
have been adjusted so that experimental data are 
described.}  
    }  }
\end{figure}

Our results for the total charm pair cross section 
in proton-proton ($pp$) collisions is shown in 
Fig.~\ref{fig:tcharm} as function of center of mass energy. 
The left panel shows the uncertainties of both approaches by varying
quark mass $m_c$ and renormalization scale $\mu_R$ 
in the intervals $1.2\GeV\le m_c \le 1.8\GeV$ and
$m_c\le \mu_R\le 2m_c$, respectively. 
The factorization scale is kept fixed at
$\mu_F=2m_c$, because in our opinion, the charm quark mass is too low for 
DGLAP evolution. 
A large fraction of the
resulting uncertainty originates 
from different possible choices of the charm
quark mass, since the total cross section behaves approximately 
like $\sigma_{\rm tot}\propto m_Q^{-2}$. 

Note that the mean value of $x_2$ increases with decreasing energy. At 
$\sqrt{s}=130\GeV$ one has $x_2\sim0.01$. For lower energies, our calculation 
is an extrapolation of the saturation model. For the highest fixed target 
energies of $\sqrt{s}\approx 40\GeV$, values of $x_2\sim0.1$ become important.
Unlike in the Drell-Yan case, which was studied in \cite{rpn}, the
dipole approach to heavy quark production does not show any unphysical 
behavior when extrapolated to larger~$x_2$. One reason for this is that
the new saturation model \cite{bgbk} assumes a realistic behavior of the
gluon density at large $x_2$. In addition, even at energies as low as
$\sqrt{s}=15\GeV$, the gluon-gluon fusion process is the 
dominant contribution to the cross section.
 
Because of the wide uncertainty bands, one can adjust $m_c$ and
$\mu_R$ in both approaches
so that experimental data are reproduced.
Then, dipole
approach and NLO parton model yield almost identical results.
However, the predictive power of the theory is rather small.
In 
Fig.~\ref{fig:tcharm} (right), we used $m_c=1.2\GeV$ and $\mu_R=1.5m_c$
for the NLO parton model calculation and $m_c=1.4\GeV$, $\mu_R=m_c$
in the dipole approach. The data points tend to
lie at the upper edge of
the uncertainty bands, so that rather small values of $m_c$ are needed
to describe them.

There are remaining uncertainties which are not shown
in Fig.~\ref{fig:tcharm} (right), because different combinations 
of $m_c$ and $\mu_R$ can also yield a good description of the data. In
addition, different PDFs will lead to different values of the
cross section at high energies, since the heavy quark cross section
is very sensitive to the low-$x$ gluon distribution. In \cite{rvt},
it was found that an uncertainty of a factor of $\sim 2.3$ remains at
$\sqrt{s}=14$ TeV (in the NLO parton model), 
even after all free parameters had been fixed to
describe total cross section data at lower energies.
In the dipole approach, 
the low-$x$ gluon distribution of the target is modeled by the dipole cross
section. Since there are not many successful parameterizations of
$\sigma_{q\bar q}$, it is difficult to quantify the uncertainty resulting
from this quantity. Using the old saturation model \cite{Wuesthoff1} instead
of the DGLAP improved one \cite{bgbk} leads only to small differences for
open charm production (this is not the case for bottom production, see below).
However, it is reasonable to expect that the uncertainty
will be at least as large as in the parton model, {\em i.e.} a factor of
2.3. Therefore, it will probably be impossible to find any signs of 
saturation in the total cross section for open charm production.
It is however interesting to see that 20 -- 30\% of the total $pp$
cross section at LHC ($\sqrt{s}=14$ TeV) goes into open 
charm \footnote{The Donnachie-Landshoff parameterization of the
total $pp$ cross section \cite{DL} predicts 
$\sigma_{\rm tot}^{pp}(\sqrt{s}=14\TeV)=100$~mb.}.

A few remarks are in order concerning the data shown in Fig.~\ref{fig:tcharm}.
The fixed target data \cite{ft}
({\em i.e} all points except the one from PHENIX \cite{PHENIX}) 
were taken in proton-nucleon collisions, and the 
quantity that was actually measured was the $D$-meson cross section 
for Feynman $x_F>0$. We corrected the data for the contribution
to open charm from charmed baryons and for the partial $x_F$ coverage
according to the prescription of \cite{report97}. 
The PHENIX point was measured in $AuAu$ collisions, assuming
that there is no
contribution to the single electron signal
from sources unrelated to charm production, such as thermal direct
photons or thermal di-leptons. The data 
point shown in the figure is for central
collisions. We did not 
correct for nuclear (anti-)shadowing effects, either.

\begin{figure}[t]
  \scalebox{0.5}{\includegraphics{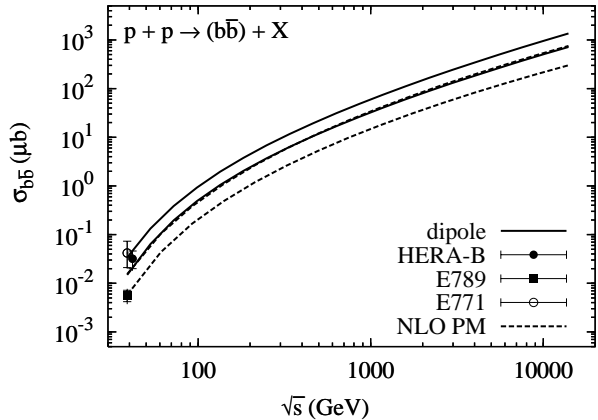}}\hfill
  \raise0.0cm\hbox{\parbox[b]{2.44in}{
   \caption{\label{fig:btotal}\em
Uncertainties of
  open $b\bar b$ pair production calculated in the dipole approach (solid)
and in the NLO parton model (dashed). 
The dipole approach seems to provide a better 
description of the data, even though HERA-B energy is too low for the dipole
approach.
  	} 
  }
}
\end{figure}

Next, we calculate the total $b\bar b$-pair cross section as function of
center of mass energy, see Fig.~\ref{fig:btotal}. 
In order to quantify the theoretical uncertainties, we vary the free
parameters over the ranges $4.5\GeV\le m_b\le 5\GeV$ and
$m_b\le\mu_R,\mu_F\le 2m_b$
Because of the large $b$-quark mass, 
uncertainties are much smaller than for open charm production. One can see 
that the dipole approach tends to predict higher values than the NLO parton
model, even though the energy dependence expected in both approaches is very
similar. In fact, the results calculated in the dipole approach with
$m_b=5\GeV$ agree almost exactly with the NLO parton model 
calculation with $m_b=4.5\GeV$. For all other 
values of $m_b$, the uncertainty bands of
the two approaches do not overlap, in contrast to
the case for open charm production.

Three measurements of
open $b\bar b$ production
are published in the literature \cite{e789,e771,herab}. 
The two values of the open bottom cross section  
measured at Fermilab \cite{e789,e771} at {\em cm.} energy
$\sqrt{s}=38.8\GeV$
differ by almost three standard deviations. 
The HERA-B measurement at slightly larger {\em cm.} energy
$\sqrt{s}=41.6\GeV$ \cite{herab} is consistent with 
the E771 \cite{e771} value.
These two points seem to be better described by the dipole approach,
though the NLO parton model (with $m_b=4.5\GeV$)
still touches the HERA-B error bar.
Note that also a different set of PDFs would not significantly
pull up the parton model curve \cite{rvt}, as a lower value of the
$b$-quark mass would do. 
With a resummation of terms from higher order corrections \cite{thresh}, 
however, the parton model can reproduce each of the three 
measurements within theoretical uncertainties, see \cite{herab}. 
On the other hand, typical values of $x_2$ which are important for $b\bar b$
production at HERA-B energy are of order $x_2\sim 0.2$, while
the parameterization
\cite{bgbk} of the dipole cross section is constrained only by DIS data
with $x_{Bj}\le 0.01$.

We point out that it is essential to use the DGLAP improved 
saturation model of \cite{bgbk} in order to obtain the same energy dependence
in the dipole approach
as in the NLO parton model. The older
saturation model of \cite{Wuesthoff1} predicts a much weaker energy 
dependence.

\begin{figure}[t]
  \centerline{\scalebox{0.5}{\includegraphics{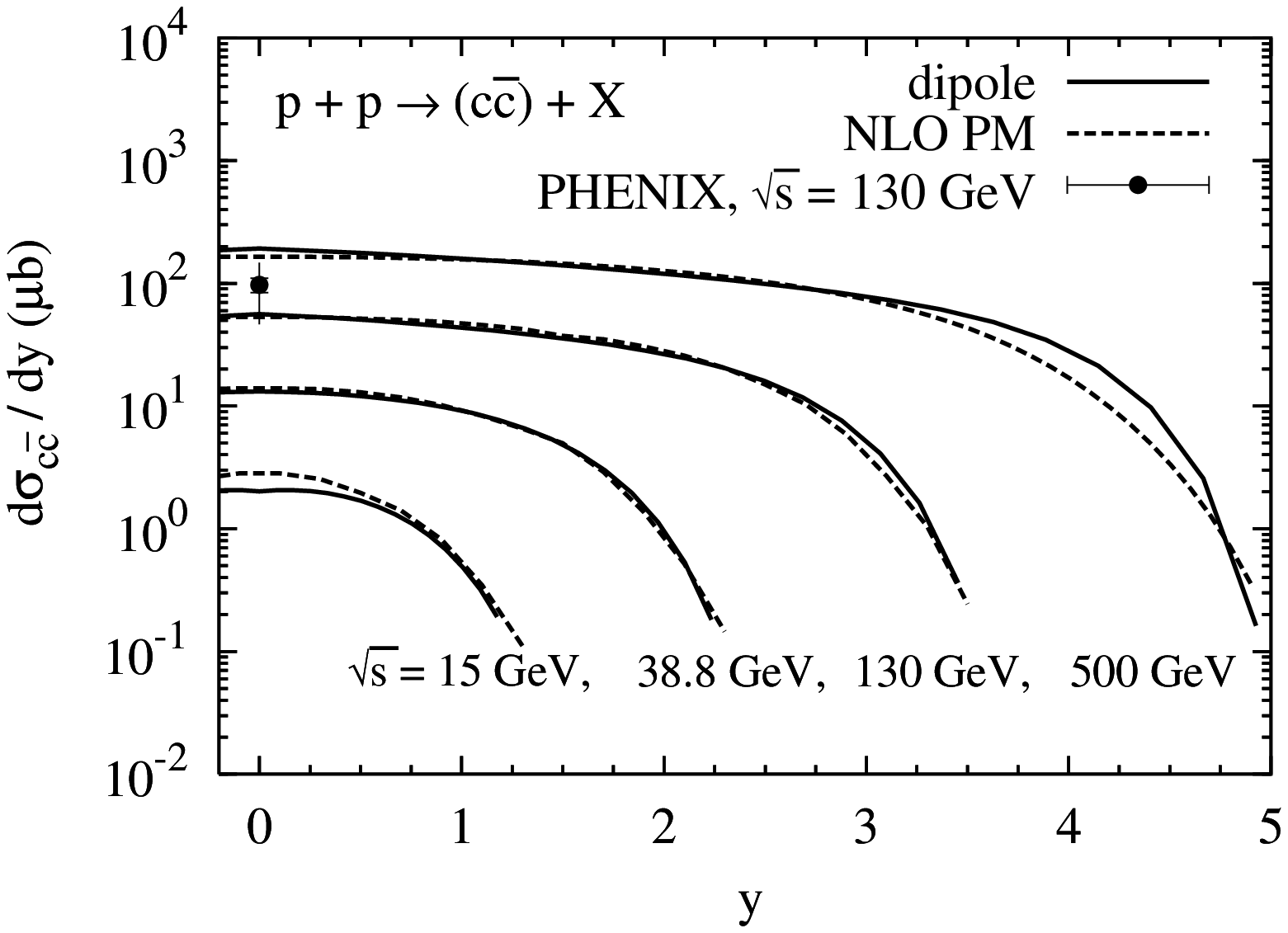}}
	      \scalebox{0.5}{\includegraphics{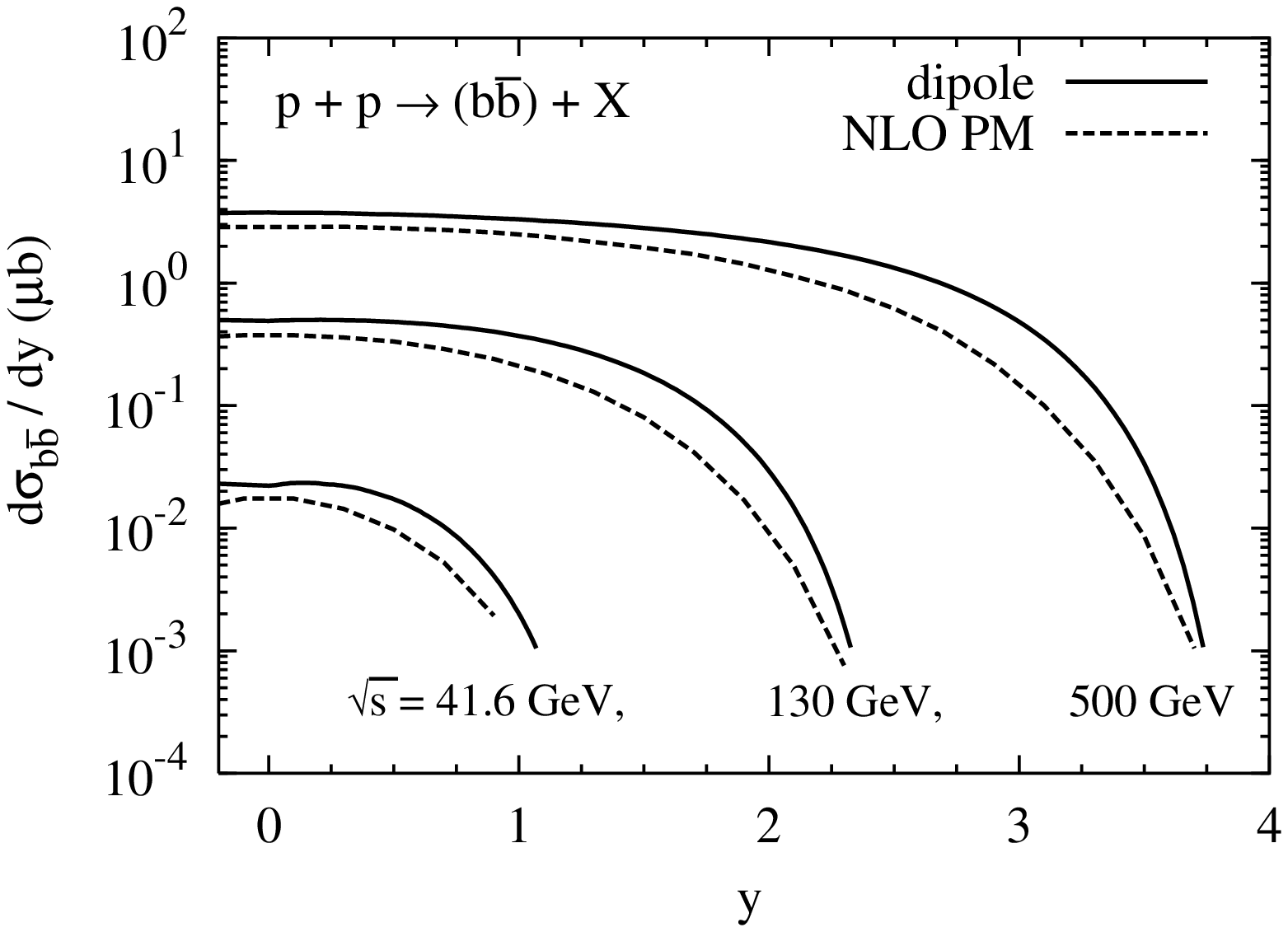}}}
\center{\parbox[thb]{13cm}{\caption{\em
      \label{fig:dy}
Rapidity distribution of heavy quark pairs. In the case of open charm (left)
free parameters of both approaches
were chosen such that total cross section data are described
i.e.\ as in Fig.~\ref{fig:tcharm} (left). The same was done for
$b\bar b$ production (right)
in the dipole approach (solid curves). In the NLO parton model
(dashed curves), we used the parameter set that yielded the largest 
cross section.
}  
    }  }
\end{figure}

Finally, we calculate the rapidity distribution of heavy quark pairs,
Fig.~\ref{fig:dy}. For open charm production, we use the values of
$m_c$, $\mu_F$ and $\mu_R$ that describe the total cross section data.
With this choice, the rapidity distributions expected in both approaches
are very similar in shape and absolute normalization. The single PHENIX
point has large errorbars, but is nevertheless well described  
by both approaches. 
The curves for $b\bar b$ production (Fig.~\ref{fig:dy}) in the dipole approach
are calculated with $m_b=\mu_F=\mu_R=4.9\GeV$, because this set of 
parameter values describes well the HERA-B point. The parton model
calculation was performed with $m_b=\mu_R=4.5\GeV$ and 
$\mu_F=2m_b$. This combination yields the upper parton model curve in 
Fig.~\ref{fig:btotal}. Even though the shape of the rapidity distributions
in the two approaches are similar, somewhat larger
cross sections are expected in the dipole approach. We have checked that
the rapidity integral over the parton model curves in Fig.~\ref{fig:dy}
reproduces the total cross sections shown in Figs.~\ref{fig:tcharm} and
\ref{fig:btotal}. 

While it is an advantage of the dipole formulation
to provide very simple formulas that allow one to absorb much of the higher
order corrections into a phenomenological parameterization of
$\sigma_{q\bar q}(x_2,\rho)$, one cannot clarify the origin of the 
discrepancy in 
normalizations without a systematic calculation of higher orders
in this approach. Such a calculation is in principle possible, but is
beyond the scope of this paper.

\section{Summary and outlook}\label{sec:sum}

In this paper, we employed the color dipole approach to study  
heavy quark pair
production in $pp$ collisions. We analytically related the leading order
dipole approach to the lowest order parton model, thereby putting the
dipole formulation on a more firm theoretical basis. In phenomenological
applications to the total heavy quark pair cross section and the pair rapidity
distribution, large theoretical uncertainties arise, mostly from different
possible choices of the heavy quark mass. This is especially true for
open charm production. Nevertheless, experimental data can be described
in the dipole approach with realistic parameter values. In kinematical
regions where no data are available, predictions from the dipole approach and
the NLO parton model for open charm production agree well after all free
parameters are fixed to describe the existing data. Note that we employ
a phenomenological parameterization of the dipole cross section, which
is supposed to include higher order effects as well, so that 
there is no justification for introducing 
an arbitrary overall normalization factor (``$K$-factor'').

Theoretical uncertainties in the cross section value,
which arise from the heavy quark mass are much
smaller in the case of open $b$-production. In this case, predictions 
from the dipole approach tend to be higher than the NLO parton model,
but the curves are similar in shape (provided QCD evolution
is included in the dipole approach). Even though two out of three 
fixed target measurements
of $b\bar b$-production seem to be better described by the dipole approach,
one has to bear in mind that
fixed target energies are 
too low for a reliable application 
of the dipole approach. Future experimental data at higher energies will
show, whether the dipole approach can reproduce the correct normalization
of the cross section.

A special advantage of the dipole formulation is the simplicity of its
formulae, which allow one to calculate the rapidity distribution
of heavy quark pairs
in only a few seconds. In addition, 
it is particularly easy to calculate nuclear effects in heavy quark
production within
this formulation \cite{npz,kt}. In fact, the latter point 
has been the original
motivation for developing this approach.

In the future, it will be necessary to 
systematically 
calculate higher order correction in the dipole approach, in
order to find out which part of these corrections can be taken into
account by a phenomenological parameterization of the dipole cross section.
This is especially important in view of the transverse momentum 
distribution of heavy quark pairs.
In the dipole approach as presented in this paper, all transverse momentum
of the pair can originate only from the intrinsic transverse
momentum of the target gluon, which is encoded in the dipole cross section.
This intrinsic transverse momentum should not be confused with the primordial
$p_T$ which is sometimes introduced in phenomenological approaches. Part of
the intrinsic $p_T$ parameterized in the dipole cross section originates
from higher orders in perturbation theory.
An investigation of higher order corrections 
will also help clarifying the origin of the
different absolute normalizations of dipole approach and parton model
for open $b$-production. 
In addition, a good theoretical understanding of the transverse momentum 
dependence of heavy quark production in the dipole formulation 
should be achieved
before one applies this approach to describe Tevatron measurements.  
We finally mention that single inclusive 
hadroproduction
of heavy quarks can be formulated in the dipole approach as well. 
We shall address these topics in a future publication.

\bigskip
{\bf Acknowledgments:}
J.R.\ is indebted to Boris Kopeliovich for 
valuable discussion.
This work was supported by
the U.S.~Department of Energy at Los Alamos
National Laboratory under Contract No.~W-7405-ENG-38.


\begin{thebibliography}{99}

\def\bibNP#1#2#3{ Nucl.\ Phys.\     {\bf #1} (#2) #3}
\def\bibZP#1#2#3{ Z.\ Phys.\        {\bf #1} (#2) #3}
\def\bibPL#1#2#3{ Phys.\ Lett.\     {\bf #1} (#2) #3}
\def\bibPR#1#2#3{ Phys.\ Rev.\      {\bf #1} (#2) #3}
\def\bibPRL#1#2#3{Phys.\ Rev.\ Lett.\ {\bf #1} (#2) #3}

\bibitem{pm1}
P.~Nason, S.~Dawson and R.~K.~Ellis,
Nucl.\ Phys.\ B {\bf 303}, 607 (1988).

\bibitem{pm2}
P.~Nason, S.~Dawson and R.~K.~Ellis,
Nucl.\ Phys.\ B {\bf 327}, 49 (1989)
[Erratum-ibid.\ B {\bf 335}, 260 (1990)].

\bibitem{pm3}
M.~L.~Mangano, P.~Nason and G.~Ridolfi,
Nucl.\ Phys.\ B {\bf 373}, 295 (1992).
A FORTRAN code for the NLO calculation is available at
http://n.home.cern.ch/n/nason/www/hvqlib.html.

\bibitem{boris}
B.~Z.~Kopeliovich, proc.\ of the workshop 
{\em Dynamical Properties of Hadrons in Nuclear Matter,}
Hirschegg, January 16 -- 21, 1995, ed.\ by H.~Feldmeyer
and W.~N\"orenberg, Darmstadt, 1995, p.~102 (hep-ph/9609385);\\
S.~J.~Brodsky, A.~Hebecker and E.~Quack,
Phys.\ Rev.\ D {\bf 55}, 2584 (1997)
[arXiv:hep-ph/9609384].

\bibitem{npz}
N.~N.~Nikolaev, G.~Piller and B.~G.~Zakharov,
J.\ Exp.\ Theor.\ Phys.\  {\bf 81} (1995) 851
[Zh.\ Eksp.\ Teor.\ Fiz.\  {\bf 108} (1995) 1554]
[arXiv:hep-ph/9412344];
Z.\ Phys.\ A {\bf 354}, 99 (1996)
[arXiv:hep-ph/9511384].

\bibitem{kt}
B.~Z.~Kopeliovich and A.~V.~Tarasov,
Nucl.\ Phys.\ A {\bf 710}, 180 (2002)
[arXiv:hep-ph/0205151].



\bibitem{gBFKL}
N.~N.~Nikolaev and B.~G.~Zakharov,
J.\ Exp.\ Theor.\ Phys.\  {\bf 78}, 598 (1994)
[Zh.\ Eksp.\ Teor.\ Fiz.\  {\bf 105}, 1117 (1994)].

\bibitem{gmrv}
S.~Gavin, P.~L.~McGaughey, P.~V.~Ruuskanen and R.~Vogt,
Phys.\ Rev.\ C {\bf 54}, 2606 (1996).

\bibitem{ev}
K.~J.~Eskola, V.~J.~Kolhinen and R.~Vogt,
Nucl.\ Phys.\ A {\bf 696}, 729 (2001)
[arXiv:hep-ph/0104124].

\bibitem{glr}
L.~V.~Gribov, E.~M.~Levin and M.~G.~Ryskin,
Nucl.\ Phys.\ B {\bf 188}, 555 (1981);
Phys.\ Rept.\  {\bf 100}, 1 (1983);\\
A.~H.~Mueller and J.~w.~Qiu,
Nucl.\ Phys.\ B {\bf 268}, 427 (1986).

\bibitem{rpn}
J.~Raufeisen, J.~C.~Peng and G.~C.~Nayak,
Phys.\ Rev.\ D {\bf 66}, 034024 (2002)
[arXiv:hep-ph/0204095].

\bibitem{borisold}
A.~B.~Zamolodchikov, B.~Z.~Kopeliovich and L.~I.~Lapidus,
JETP Lett.\  {\bf 33}, 595 (1981)
[Pisma Zh.\ Eksp.\ Teor.\ Fiz.\  {\bf 33}, 612 (1981)].

\bibitem{mp}
H.~I.~Miettinen and J.~Pumplin,
Phys.\ Rev.\ D {\bf 18}, 1696 (1978).

\bibitem{ct}
G.~Bertsch, S.~J.~Brodsky, A.~S.~Goldhaber and J.~F.~Gunion,
Phys.\ Rev.\ Lett.\  {\bf 47}, 297 (1981);\\
S.~J.~Brodsky and A.~H.~Mueller,
Phys.\ Lett.\ B {\bf 206}, 685 (1988).

\bibitem{fsdipole} 
B.~Blaettel, G.~Baym, L.~L.~Frankfurt and M.~Strikman,
Phys.\ Rev.\ Lett.\  {\bf 70}, 896 (1993);\\
L.~L.~Frankfurt, A. Radyushkin and M.~Strikman,
Phys.\ Rev.\ D {\bf 55}, 98 (1997)
[hep-ph/9610274].

\bibitem{nlo}
V.~S.~Fadin and L.~N.~Lipatov,
Phys.\ Lett.\ B {\bf 429}, 127 (1998)
[arXiv:hep-ph/9802290];\\
M.~Ciafaloni and G.~Camici,
Phys.\ Lett.\ B {\bf 430}, 349 (1998)
[arXiv:hep-ph/9803389].

\bibitem{bgbk}
J.~Bartels, K.~Golec-Biernat and H.~Kowalski,
Phys.\ Rev.\ D {\bf 66}, 014001 (2002)
[arXiv:hep-ph/0203258].

\bibitem{Wuesthoff1} 
K.~Golec-Biernat and M.~W\"usthoff,
Phys.\ Rev.\ D {\bf 59}, 014017 (1999)
[hep-ph/9807513];
Phys.\ Rev.\ D {\bf 60}, 114023 (1999)
[hep-ph/9903358].

\bibitem{grv}
M.~Gl\"uck, E.~Reya and A.~Vogt,
Eur.\ Phys.\ J.\ C {\bf 5}, 461 (1998)
[arXiv:hep-ph/9806404].

\bibitem{cernlib} H.~Plothow-Besch, 
Int.\ J.\ Mod.\ Phys.\ A {\bf 10}, 2901 (1995);
``PDFLIB: Proton, Pion and Photon Parton Density Functions, Parton Density 
Functions of the Nucleus and $\alpha_s$ Calculations'', 
User's Manual - Version 8.04, W5051 PDFLIB, 2000.04.17, CERN-PPE.

\bibitem{rvt}
R.~Vogt,
arXiv:hep-ph/0203151.

\bibitem{DL}
A.~Donnachie and P.~V.~Landshoff,
Phys.\ Lett.\ B {\bf 296}, 227 (1992)
[arXiv:hep-ph/9209205].

\bibitem{ft}
K.~Kodama {\it et al.}  [Fermilab E653 Collaboration],
Phys.\ Lett.\ B {\bf 263}, 573 (1991);\\
R.~Ammar {\it et al.},
Phys.\ Rev.\ Lett.\  {\bf 61}, 2185 (1988);\\
M.~Aguilar-Benitez {\it et al.}  [LEBC-EHS Collaboration],
Z.\ Phys.\ C {\bf 40}, 321 (1988);\\
G.~A.~Alves {\it et al.}  [E769 Collaboration],
Phys.\ Rev.\ Lett.\  {\bf 77}, 2388 (1996)
[Erratum-ibid.\  {\bf 81}, 1537 (1998)];\\
S.~Barlag {\it et al.}  [ACCMOR Collaboration],
Z.\ Phys.\ C {\bf 39}, 451 (1988).

\bibitem{PHENIX}
K.~Adcox {\it et al.}  [PHENIX Collaboration],
Phys.\ Rev.\ Lett.\  {\bf 88}, 192303 (2002)
[arXiv:nucl-ex/0202002].

\bibitem{report97}
S.~Frixione, M.~L.~Mangano, P.~Nason and G.~Ridolfi,
Adv.\ Ser.\ Direct.\ High Energy Phys.\  {\bf 15}, 609 (1998)
[arXiv:hep-ph/9702287].

\bibitem{e789}
D.~M.~Jansen {\it et al.} [E789 Collaboration],
Phys.\ Rev.\ Lett.\  {\bf 74}, 3118 (1995).

\bibitem{e771}
T.~Alexopoulos {\it et al.}  [E771 Collaboration],
Phys.\ Rev.\ Lett.\  {\bf 82}, 41 (1999).

\bibitem{herab}
I.~Abt {\it et al.}  [HERA-B Collaboration],
arXiv:hep-ex/0205106.

\bibitem{thresh}
R.~Bonciani, S.~Catani, M.~L.~Mangano and P.~Nason,
Nucl.\ Phys.\ B {\bf 529}, 424 (1998)
[arXiv:hep-ph/9801375];\\
N.~Kidonakis, E.~Laenen, S.~Moch and R.~Vogt,
Phys.\ Rev.\ D {\bf 64}, 114001 (2001)
[arXiv:hep-ph/0105041].

\end{thebibliography}
\end{document}